\documentclass[abbrv,aps,prl,twocolumn,showpacs,preprintnumbers,amsmath,amssymb,
superscriptaddress,nofootinbib]{revtex4}

\usepackage{graphicx}
\usepackage{dcolumn}
\usepackage{bm}
\begin{document}
\title{NiO: Correlated Bandstructure of a Charge-Transfer Insulator}
\author{J. Kune\v{s}}
\email{jan.kunes@physik.uni-augsburg.de}
\affiliation{Theoretical Physics III, Center for Electronic Correlations and Magnetism, Institute of Physics, 
University of Augsburg, Augsburg 
86135, Germany}
\affiliation{Institute of Physics,
Academy of Sciences of the Czech Republic, Cukrovarnick\'a 10,
162 53 Praha 6, Czech Republic}
\author{V.~I.~Anisimov}
\affiliation{Institute of Metal Physics, Russian Academy of
Sciences-Ural Division, 620041 Yekaterinburg GSP-170, Russia}
\author{S. L. Skornyakov}
\affiliation{Ural State Technical University-UPI,
620002 Yekaterinburg, Russia}
\author{A.~V.~Lukoyanov}
\affiliation{Ural State Technical University-UPI,
620002 Yekaterinburg, Russia}
\author{D. Vollhardt}
\affiliation{Theoretical Physics III, Center for Electronic Correlations and Magnetism, Institute of Physics,
University of Augsburg, Augsburg
86135, Germany}
\date{\today}

\begin{abstract}
The bandstructure of the prototypical charge-transfer insulator NiO is computed
by using a combination of an {\it ab initio} bandstructure method and the dynamical mean-field theory
with a quantum Monte-Carlo impurity solver. Employing a Hamiltonian which includes both Ni-$d$ and O-$p$ orbitals
we find excellent agreement with the energy bands determined from angle-resolved photoemission spectroscopy. 
This solves a long-standing problem in solid state theory.
Most notably we obtain the low-energy Zhang-Rice bands with strongly ${\mathbf k}$-dependent orbital character
discussed previously in the context of low-energy model theories. 

\end{abstract}
\maketitle

The quantitative explanation of the electronic structure of transition metal oxides (TMOs) and
other materials with correlated electrons has been
a long-standing challenge in condensed matter physics. While the basic concept explaining why materials 
such as NiO are insulators was formulated by Mott already a long time ago \cite{mott49}, 
the development of an appropriate, material-specific computational scheme proved to be a formidable task. 
The electronic structure of the late TMOs, including the cuprate superconductors, is not only affected
by the electronic correlations, it is further complicated by the  
hybridization between the transition metal $d$-states and O $p$-bands located between the lower and 
upper Hubbard bands formed by the transition metal $d$ orbitals. For such materials 
Zaanen, Sawatzky and Allen \cite{zaa85} introduced the term "charge transfer insulator", 
a prototypical example of which is NiO. In principle the simple
crystal structure of NiO allows for a straightforward comparison between theory and experiment.
However, a theoretical description of the NiO bandstructure 
is made difficult by the competition between the local many-body effects, due to strong Coulomb interaction between Ni $d$ electrons, 
and the band dispersion, due to the lattice periodicity, both observed with the angle-resolved 
photoemission spectroscopy (ARPES) \cite{she90,she91}. 

In this Letter we use a combination of a conventional bandstructure approach, based on the local density
approximation (LDA), and the dynamical mean-field theory (DMFT) \cite{met89,rmp,pt} to investigate the bandstructure
of NiO. No adjustable parameters enter. While the application of the LDA+DMFT \cite{ldadmfta,ldadmftb,rmp06} framework has proven successful for the 
early TMOs,
the charge-transfer materials were routinely avoided due to the additional complexity arising from 
the presence of $p$-bands. In the present work the O $p$-orbitals and their hybridization
with Ni $d$-orbitals are explicitly included, thus allowing for a unified description 
of the full spectrum. Our results reveal a non-trivial effect of the $p-d$ hybridization
in strongly correlated system studied so far only in terms of simple models \cite{zha88,fuj84,esk91}. 

The application of the standard bandstructure theory to NiO is marked by a failure of LDA to produce an insulating
groundstate \cite{mat72}. The antiferromagnetic order within LDA \cite{ter84},
despite rendering NiO an insulator,
does not present much of an improvement since (i) the band gap is severely underestimated, (ii) the experimentally
observed lower Hubbard band is completely missing, and (iii) static quantities such as the local magnetic moment
do not agree well with experiment.
Moreover, the ARPES data of Tjernberg {\it et al.} \cite{tje96} measured across the N\'eel temperature $T_N$=525 K
show that the NiO bandstructure is rather insensitive to the magnetic order.
The first attempt to include the strong on-site correlations into the first-principles bandstructure methods
was the LDA+U theory of Anisimov {\it et al.} \cite{ani91}. 
The static, orbitally dependent self-energy of LDA+U
enforces a separation of the occupied and unoccupied $d$-bands and thus opens a gap of the experimentally observed size.
This in turn leads to a significant improvement in the description of 
groundstate properties such as the local moment or the lattice constant \cite{savrasov}.
Despite this success LDA+U does not provide a good description
of the photoemission spectra since it places almost all of the valence $d$ spectral weight 
into the lower Hubbard band.
Exact diagonalization studies of Fujimori {\it et al.} \cite{fuj84} on small clusters provide
strong evidence that the dynamical correlations are necessary to capture properly the distribution 
of the $d$ spectral weight within the valence band.

A systematic inclusion of local dynamical correlations into lattice models was made possible by 
the dynamical mean-field theory (DMFT) \cite{met89,rmp,pt}. 
The connection of DMFT with bandstructure methods, usually referred to as LDA+DMFT scheme \cite{ldadmfta,ldadmftb,rmp06}, provides access to 
material-specific single-particle spectra as well as more general correlation functions.
LDA+DMFT calculations usually start
with the construction of an effective Hamiltonian from the converged LDA bandstructure 
using a projection of the low-energy bands onto Wannier orbitals. Typically only the transition
metal $d$ bands are included in the construction, which amounts to integrating out
O $p$-states \cite{ren06} at the price of having more extended Wannier orbitals,
a procedure only justified for the early TMOs with low-lying $p$-bands.
In the present study we project Wannier states out of a single energy window including both $p$ and $d$ bands,
resulting in a more localized O-$p$ and Ni-$d$ Wannier orbitals with mutual hybridization.
\begin{figure}
\includegraphics[angle=180,width=\columnwidth,clip]{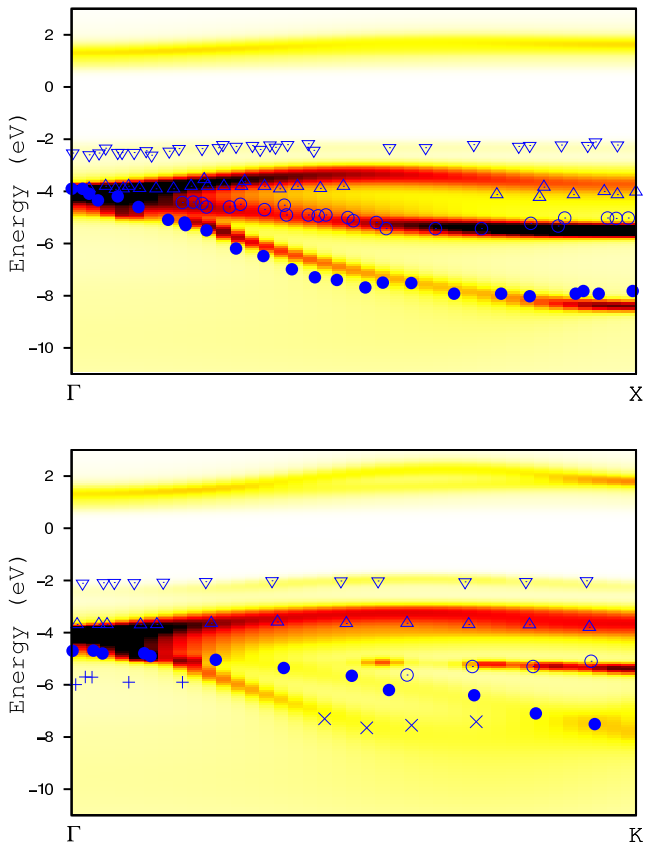}
\caption{\label{fig:gx}(color online) The ${\mathbf k}$-resolved total spectral function $A({\mathbf k},\omega)$ along
the $\Gamma$-X (upper panel) and $\Gamma$-K (lower panel) lines in the Brillouin zone depicted as a contour plot. 
The symbols represent the experimental bands of Shen {\it et al.} \cite{she91}. The theoretical gap edge
was aligned with the experimental one.}
\end{figure}

The computation proceeds in several steps: (i) the construction of the effective Hamiltonian
from converged LDA results, (ii) the self-consistent solution of the DMFT equations on the 
Matsubara contour, (iii) the analytic continuation of single-particle Green function 
to the real frequencies, (iv) the calculation of the self-energy on the real axis, and the computation
of the generalized bandstructure.
We use a projection onto Wannier functions \cite{ani05} to obtain the Hamiltonian discussed in Ref. \onlinecite{kun07}.
Next we solve the DMFT equations iteratively on the Matsubara
contour, a key part of which is the auxiliary impurity problem handled by quantum Monte-Carlo (QMC) method \cite{qmc}.  
To ensure ergodic sampling we introduced global moves between ferro- and anti-ferromagnetic
configurations of local $e_g$ spins. The computational parameters are the same as in Ref. \onlinecite{kun07}.
The $d$-spectral functions were obtained by the maximum entropy method \cite{mem} applied separately 
for the $e_g$ and $t_{2g}$ symmetry. 
The local self-energy $\Sigma(\omega^+)$, which is formally an $8\times8$ matrix with the only non-zero elements
on the diagonal of $dd$ block, is obtained by solving the equation
\begin{equation}
G_{dd}(\omega^+)=\sum_{\mathbf{k}}\large(\omega^++\mu-h_{\mathbf{k}}-\Sigma(\omega^+)\large)^{-1}_{dd},
\end{equation}
simultaneously for $e_g$ and $t_{2g}$ symmetry. 
Here the $G_{dd}$ is the diagonal element of the Green function corresponding to either $e_g$ or $t_{2g}$ symmetry, 
$h_{\mathbf{k}}$ is the $8\times8$ Hamiltonian
matrix on a mesh of ${\mathbf k}$-points and $\mu$ is the self-consistently determined chemical potential.
The equation is solved approximately on the contour $\Im\omega^+=0.05eV$ subject
to the constraints $\Im\Sigma(\omega^+)<0$ and Kramers-Kronig relations, which are, however, satisfied automatically
for good quality QMC data.

In Figs. \ref{fig:gx} we compare the theoretical bands, represented by the
${\mathbf k}$-dependent spectral density $A({\mathbf k},\omega)$, along the $\Gamma-X$ and
$\Gamma-K$ lines in the Brillouin zone with ARPES data of Ref. \onlinecite{she91}. Both theory and experiment exhibit two 
relatively flat bands at -2 and -4 eV followed by several dispersive bands in the -4 to -8 eV range and a broad incoherent peak
around -10 eV. Overall we find an excellent agreement. 
The deviations around the $\Gamma$ point in the lower panel of Fig. \ref{fig:gx} are due to the inaccuracy 
in the location of the $\Gamma$ point in off-normal-emission experiment \cite{she91}. 
The crosses near the $\Gamma$ point mark a weak band which was interpreted as a consequence
of AFM order \cite{she91} and is therefore not expected to be found in the paramagnetic phase investigated here.  
\begin{figure}
\includegraphics[angle=270,width=\columnwidth,clip]{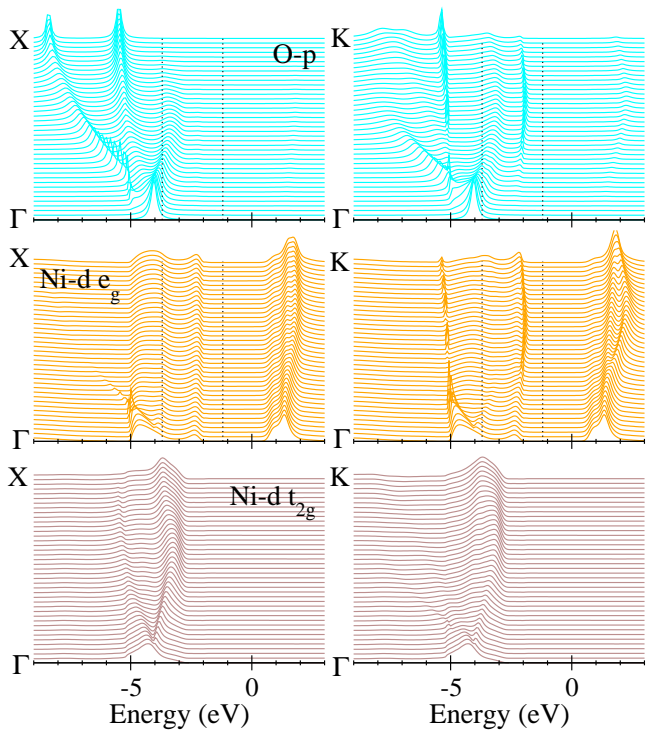}
\caption{\label{fig:ake-gx}(color online) The orbitally decomposed spectral function $A_{\nu\nu}({\mathbf k},\omega)$ 
along the $\Gamma$-X (left column) and $\Gamma$-K (right column) lines in the Brillouin zone
plotted as $\frac{A_{\nu\nu}({\mathbf k},\omega)}{C+A_{\nu\nu}({\mathbf k},\omega)}$. 
The panels from top to bottom show the O-$p$, Ni-$d-e_g$ and Ni-$d-t_{2g}$ contributions.
Here $C=1.5,2$ for the $p$ and $d$ projections, respectively. Detail of the uppermost valence band marked by the
dotted lines is shown in Fig. \ref{fig:zr}.}
\end{figure}
In Fig. \ref{fig:ake-gx} we show the orbital decomposition $A_{\nu\nu}({\mathbf k},\omega)$ 
of the spectral density visualized as a shape-preserving function  
$\frac{A_{\nu\nu}({\mathbf k},\omega)}{C+A_{\nu\nu}({\mathbf k},\omega)}$, in order to capture
both sharp and broader features in a single plot.

We start the discussion of our results by considering the limit of vanishing $p-d$ hybridization,
in which case
the entire $d$ spectral weight is located in more or less featureless Hubbard bands located below
the $p$-band manifold \cite{kun07} and the holes in the uncorrelated $p$-bands have infinite lifetimes.
The $p-d$ hybridization changes this picture qualitatively. In particular,
two additional bands of mixed character appear at -2 and -4 eV of Fig. \ref{fig:gx}.
These bands contain about half of the valence $d$ spectral weight as can be seen
from the ${\mathbf k}$-integrated spectrum of Ref. \onlinecite{kun07}.
In addition, the lower Hubbard band is broadened  due to the opening of the $p-d$ decay channel
for the $d$-holes introducing a pronounced asymmetry between the upper and lower Hubbard bands 
\footnote{Compare to the symmetric upper and lower Hubbard bands of $e_g$ symmetry in Fig. 3 of Ref. \onlinecite{ren06}, where
$p-d$ hybridization was not included.}
(see the middle panel of Fig. \ref{fig:ake-gx}). 
Reciprocally, the $p$ bands possess a finite ${\mathbf k}$-dependent width due to the coupling to the correlated
$d$ bands.
While full quantitative comparison to the ARPES data is not possible because
of missing dipole matrix elements, the ${\mathbf k}$-dependent broadening of the $p$ bands (see the
upper panel of Fig. \ref{fig:ake-gx}) agrees well with the experimental observations \cite{she90,she91}. 
In particular, we point out the broadening of the otherwise sharp $p$ bands
observed around the midpoint of the  $\Gamma$-X line and a considerable smearing of the 
lowest two $p$ bands near the K point (upper right panel of Fig. \ref{fig:ake-gx}),
which are hardly recognizable in the experimental spectra.
\begin{figure}
\includegraphics[angle=270,width=\columnwidth,clip]{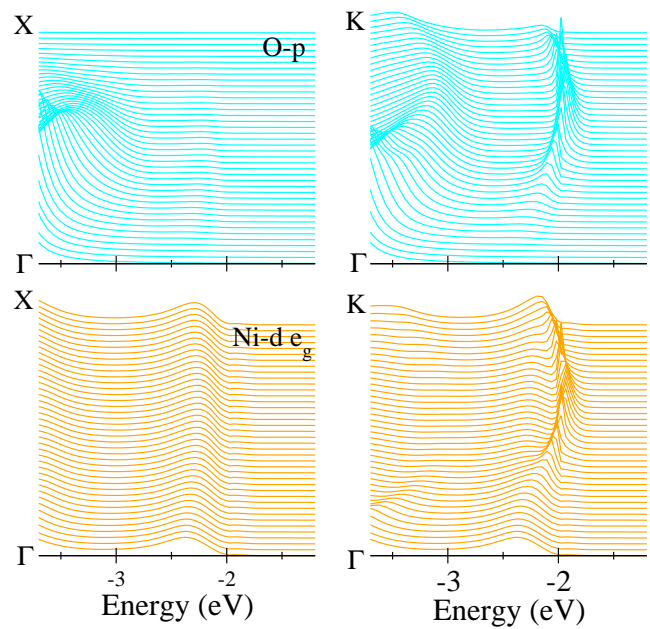}
\caption{\label{fig:zr}(color online) Detail of the uppermost valence band along the $\Gamma$-K (right) and
$\Gamma$-X (left) lines. The top panels show the O-$p$ contribution $A_{pp}({\mathbf k},\omega)$,
while Ni-$d$ contribution $A_{dd}({\mathbf k},\omega)$
$A_{dd}({\mathbf k},\omega)$ of the $e_g$ symmetry is shown in the bottom panels.}
\end{figure}

The low energy bands of charge-transfer systems, 
especially in the case of high temperature cuprate superconductors, have been subject of
numerous theoretical investigations. This was initiated by Zhang and Rice \cite{zha88} who constructed
an effective $t-J$ Hamiltonian for holes doped to the copper-oxygen plane and who introduced
the notion of a bound state between the $p$-hole and $d$-spin known as Zhang-Rice singlet.
Using a canonical transformation of the Hubbard model onto the spin-fermion model 
Eroles {\it et al.} \cite{ero99} found a strong ${\mathbf k}$-dependence in the orbital composition
of the Zhang-Rice band. Ba{\l}a {\it et al.} \cite{bal94, bal00} applied
a generalized spin-fermion model to a 2D slab of NiO and obtained qualitative agreement with the
corresponding part of the ARPES spectrum. 
They found very strong ${\mathbf k}$-dependence of the spectral weight in the uppermost valence band. 
To a good approximation the latter corresponds to the $p$-spectral weight
of the original multi-band Hubbard model \cite{ero99} and is thus directly comparable
to the results of the present study. 
Like Ba{\l}a {\it et al.}, we find a vanishing $p$ spectral weight in the uppermost valence
band along the $\Gamma$-X ($\langle10\rangle$ in Ref. \onlinecite{bal94}) line as well as a substantial
$p$ contribution in the $\Gamma$-K direction ($\langle11\rangle$ in Ref. \onlinecite{bal94})
shown in Fig. \ref{fig:zr}.
In addition, we find a rather ${\mathbf k}$-independent $d$ contribution to the uppermost valence band,
which was suggested in Ref. \onlinecite{bal00}
and which reflects the local character of the Zhang-Rice bound state, a doublet in the case of NiO.
Unlike model theories, which are restricted to a special part of the Hilbert space, 
the LDA+DMFT scheme provides a unified picture of all energy scales and avoids
any adjustable parameters. The good agreement with the experimental data and the low-energy model
theory found in this work demonstrates the capability of DMFT to adequately describe charge-transfer systems 
and puts the earlier model results on a solid foundation.

In conclusion, by employing dynamical mean-field theory combined with LDA electronic structure calculation
we presented the solution to a long-standing problem, the computation of the full 
valence bandstructure of a charge-transfer insulator. 
We obtained a very good agreement with the ARPES data of Shen {\it et al.} \cite{she90,she91} without any adjustable parameters.
We found well-separated Zhang-Rice bands at the top valence manifold with strongly ${\mathbf k}$-dependent orbital
composition. 
Our results clearly demonstrate the capability of DMFT to treat, upon explicit inclusion of
$p-d$ hybridization, the late transition-metal oxides and charge-transfer systems in general. 

J.K. gratefully acknowledges the Research Fellowship of the Alexander von Humboldt Foundation.
This work was supported by the SFB 484 of the Deutsche Forschungsgemeinschaft (J.K., D.V.),
by the Russian Foundation for Basic Research
under the grants RFFI-06-02-81017, RFFI-07-02-00041 (V.I.A., S.L.S. and A.V.L.) and by
the Dynasty Foundation (A.V.L.).

\end{document}